# Implementation of Program Part at Automated Workplace for a Teaching Department

Vitaly Ushakov
Saint-Petersburg State University of Aerospace Instrumentation
Saint-Petersburg, Russia
mr.vitaly.ushakov@yandex.ru

***Abstract*—The problem of improving the efficiency of the teaching department through the development of teaching department work area is described. Development of an automated workplace of a teaching department who allows to realize monitoring of progress of students, monitoring of mastering of disciplines by students, is synchronized with an automated workplace of the teacher of the higher school and autocompletes the report of movement of the contingent. Besides, the designed system allows to increase efficiency and efficiency of activities of employees of a teaching department.**

## I. Introduction

Scientifically technical progress has led to that practically everyone uses electronic computers in the activity, whether it be the normal stationary computer, a laptop or a pad.

Besides, now in Russia the great attention at the state level is given to society information as a whole and education spheres in particular. The federal, interdepartmental and branch programs directed on the decision of actual tasks of information of education, including development of an infrastructure of educational information field, development of electronic educational resources, improvement of professional skill of teachers in application field information and communication technologies, their implementation in the organization of educational process are implemented.

Increase of management efficiency by a teaching department of a higher educational institution is one of the key tasks facing a manual of a teaching department. The constant magnification of volumes and intensity of information streams leads to necessity of usage of information means and technologies for increase of efficiency of handling, reliability and convenience of representation. But for the full changeover of traditional paper document circulation of a teaching department it is necessary to provide toolkit for display in the electronic form as a minimum of the same functions, however a part of functions thus is necessary for automating.

For the decision of this task the automated workplace of a teaching department has been developed. Our application allows [1]:

- to simplify operation of employees of a teaching department,
- to reduce an amount of paper documents,
- to eliminate financial expenses for printing of documents,
- to lower probability of a human error,
- training activity formalization.

Also helps to automate the following functions:

- calculation of the academic debts,
- filling of the report of movement of the contingent,
- control of development of subject matters,
- monitoring of progress of students/

The automated workplace represents set of the hardware-software means providing interaction of the person from a computer, gives possibility of information input and its output. In the given report I will consider software of an automated workplace of a teaching department, because hardware will be considered in other report.

## II. About an Automated Workplace of a Teaching Department

The automated workplace of a teaching department represents information system electronic magazine containing information on being trained students, the deducted students and the students being in the academic holiday. The application contains information necessary for formation of the monthly report of movement of the contingent of students. And as it contains information on dates of delivery by students of examinations, offsets and term papers. Appearance of the main application window is shown on Fig. 1 [2].

This application uses some tables of basic data [3]:

- single information. The basic data entered once concern her: numbers of groups, biographical particulars about students etc.
- periodic information. It contains curricula on a semester, information from orders on staff of students etc.

- operational information. Information on total certification concerns her from sheets and record books.

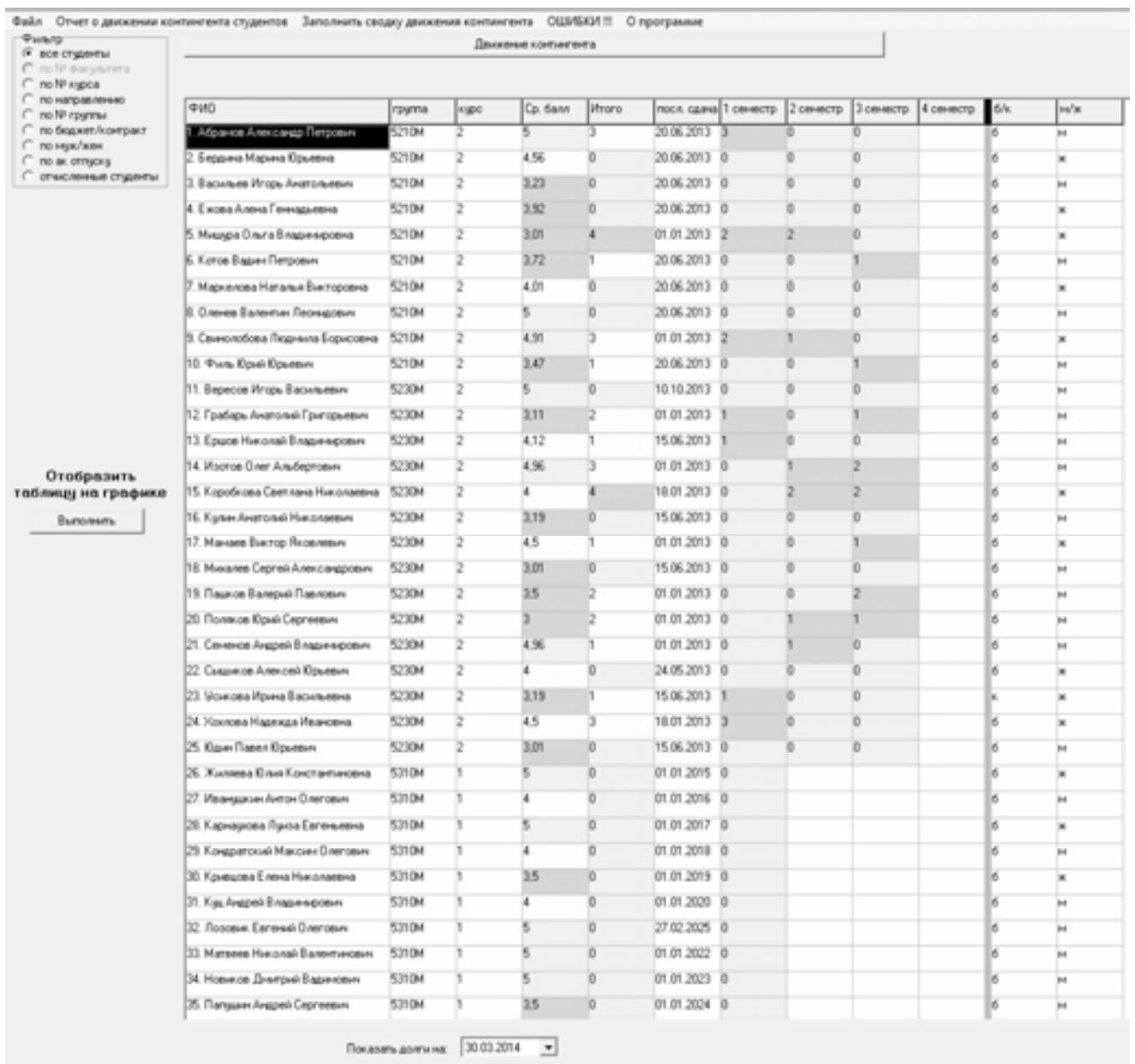

Fig. 1. Type of the main window

In the main window it is possible to choose type filtration students:

- all students (without students on the academic holiday and deducted students),
- according to course number,
- in the direction,
- according to group number. It is shown on Fig. 2.
- on form of education (budget/contract),
- on a floor (male/female),
- on the academic holiday. It is shown on Fig. 3.
- the deducted students. All information is available only in browse mode. It is shown on Fig. 4.

Besides, it is possible, to sort the summary table by any column, to edit dates of deposit disciplines and personal information on students, to display the summary table on graphics, to execute movement of the contingent and to show the academic debts for the chosen date in the past.

| ФИО | группа | курс | Ср. балл | Итого | посл. сдача | 1 семестр | 2 семестр | 3 семестр | 4 семестр | б/к | м/ж |
|---|---|---|---|---|---|---|---|---|---|---|---|
| 1. Абрамов Александр Петрович | 5210M | 2 | 5 | 3 | 20.06.2013 | 3 | 0 | 0 | | 6 | м |
| 2. Бердина Марина Юрьевна | 5210M | 2 | 4,56 | 0 | 20.06.2013 | 0 | 0 | 0 | | 6 | ж |
| 3. Васильев Игорь Анатольевич | 5210M | 2 | 4,23 | 0 | 20.06.2013 | 0 | 0 | 0 | | 6 | м |
| 4. Ежова Алена Геннадьевна | 5210M | 2 | 4,92 | 0 | 20.06.2013 | 0 | 0 | 0 | | 6 | ж |
| 5. Мишура Ольга Владимировна | 5210M | 2 | 3,01 | 4 | 01.01.2013 | 2 | 2 | 0 | | 6 | ж |
| 6. Котов Вадим Петрович | 5210M | 2 | 3,72 | 1 | 20.06.2013 | 0 | 0 | 1 | | 6 | м |
| 7. Маркелова Наталья Викторовна | 5210M | 2 | 4,01 | 0 | 20.06.2013 | 0 | 0 | 0 | | 6 | ж |
| 8. Оленев Валентин Леонидович | 5210M | 2 | 5 | 0 | 20.06.2013 | 0 | 0 | 0 | | 6 | м |
| 9. Свинолобова Людмила Борисовна | 5210M | 2 | 4,91 | 3 | 01.01.2013 | 2 | 1 | 0 | | 6 | ж |
| 10. Филь Юрий Юрьевич | 5210M | 2 | 3,47 | 1 | 20.06.2013 | 0 | 0 | 1 | | 6 | м |

Fig. 2. Student's group

| ФИО | группа | курс | Ср. балл | Итого | посл. сдача | 1 семестр | 2 семестр | 3 семестр | 4 семестр | б/к | м/ж |
|---|---|---|---|---|---|---|---|---|---|---|---|
| 1. Суязова Мария Михайловна | 5330M | 1 | 0 | 3 | 26.01.2014 | 3 | | | | 6 | ж |
| 2. Мусиенко Николай Николаевич | 5210M | 2 | 3,27 | 4 | 26.01.2019 | 1 | 3 | 0 | | 6 | м |
| 3. Решетняк Людмила Михайловна | 5330M | 1 | 0 | 3 | 26.01.2014 | 3 | | | | 6 | ж |

Fig. 3. Students being in the academic holiday

| ФИО | группа | курс | Ср. балл | б/к | м/ж | дата | причина | "долги" |
|---|---|---|---|---|---|---|---|---|
| 1. Коломойцев Владимир Сергеевич | 5210M | 2 | 0 | 6 | м | 24.01.2014 | болезнь | 2 |

Fig. 4. The deducted students

The application is written to Delphi 7 in the Object Pascal language [4-7]. For prevention mistakes at a development stage and testing program the log was used [8].All this information is stored in a database of the program which in turn takes place in two Microsoft Excel files[9-12]: one file with data on students (a surname, a name, a floor, a course, the budget/contract), another - with curricula of groups. Storage of information is organized by a principle: on one sheet of the workbook information on one group is placed. You can see files with program data at Fig. 5 and Fig. 6.

| | 1 | 2 | 3 | 4 | 5 | 6 | 7 | 8 | 9 | 10 | 11 | 12 | 13 |
|---|---|---|---|---|---|---|---|---|---|---|---|---|---|
| 1 | Фамилия | Имя | Отчество | Курс | Б/К | Пол | № студ. | Ср. балл | | | Даты сдач экзаменов | | |
| 2 | Абрамо | Алексан | Петрови | 2 | Бюджет | Мужской | 012/398 | 5 | | | 12.01.2013 | | 04.01.201 |
| 3 | Бердина | Марина | Юрьевн | 2 | Бюджет | Женский | 012/598 | 4,56 | | | 12.01.201 | 12.01.201 | 04.01.201 |
| 4 | Василье | Игорь | Анатоль | 2 | Бюджет | Мужской | 012/798 | 4,23 | | | 12.01.201 | 12.01.201 | 04.01.201 |
| 5 | Ежова | Алена | Геннадь | 2 | Бюджет | Женский | 012/898 | 4,92 | | | 12.01.201 | 12.01.201 | 04.01.201 |
| 6 | Мишура | Ольга | Владими | 2 | Бюджет | Женский | 012/000 | 3,01 | | | 01.01.201 | 01.01.201 | 01.01.2013 |
| 7 | Котов | Вадим | Петрови | 2 | Бюджет | Мужской | 012/300 | 3,72 | | | 12.01.201 | 12.01.201 | 04.01.201 |
| 8 | Маркел | Наталья | Викторо | 2 | Бюджет | Женский | 012/300 | 4,01 | | | 12.01.201 | 12.01.201 | 04.01.201 |
| 9 | Оленев | Валенти | Леонидс | 2 | Бюджет | Мужской | 012/300 | 5 | | | 12.01.201 | 12.01.201 | 04.01.201 |
| 10 | Свинопс | Людмил | Борисов | 2 | Бюджет | Женский | 011/978 | 4,91 | | | | 01.01.201 | 01.01.201 |
| 11 | Филь | Юрий | Юрьеви | 2 | Бюджет | Мужской | 012/501 | 3,47 | | | 12.01.201 | 12.01.201 | 04.01.201 |

Fig. 5. Group.xls – data about students

Our application does not contain estimates, and contains only dates of delivery of disciplines because for definition amount of the academic debts there is enough date. Besides, at assignment it is more important to know date of the last delivery, rather than an assessment of the student.

| | 1 | 2 | 3 |
|---|---|---|---|
| 1 | Номер семестра | Название предмета | Вид контроля |
| 2 | 1 | НИР в семестре | 6 |
| 3 | 1 | Научно-технический семинар | 2 |
| 4 | 1 | Теория электромагнитной совместимости радиоэлектронных средств и систем | 1 |
| 5 | 1 | Математические основы криптологии | 1 |
| 6 | 1 | Компьютерные технологии в науке и телекоммуникациях | 2 |
| 7 | 1 | Методы моделирования и оптимизации | 1 |
| 8 | 1 | Философские и психологические проблемы творчества | 1 |
| 9 | 1 | Современные проблемы информатики и вычислительной техники | 2 |
| 10 | 1 | Технология разработки программного обеспечения | 2 |
| 11 | 1 | Методы оптимизации | 1 |
| 12 | 1 | Теория принятия решений | 2 |
| 13 | 1 | Сети ЭВМ и телекоммуникации | 1 |
| 14 | 1 | Научно-исследовательская работа в семестре | 6 |
| 15 | 1 | Вычислительные системы | 1 |
| 16 | 1 | Проектирование встроенных систем на СБИС | 1 |
| 17 | 2 | Интерфейсы и протоколы инфокоммуникационных систем | 1 |
| 18 | 2 | Компьютерные технологии в науке и телекоммуникациях | 1 |
| 19 | 2 | Психология и педагогика (высшей школы) | 2 |
| 20 | 2 | Иностранный язык | 1 |

Fig. 6. Training_plan.xls – data about training plan

## III. MONITORING PROGRESS STUDENTS

*A. Monitoring progress*

The automated workplace of a teaching department allows to carry out monitoring of progress of students.

Monitoring of progress of students is carried out for the purpose of obtaining of the necessary information on performance of the schedule of educational process by them, levels of achievement of an object in view of training, stimulation of independent operation of students. It assists improving of the organisation and carrying out of studies, and also gain of responsibility of students for quality of the study in university. For this purpose in a SUAI it is developed and position about modularly-rating system of an estimation of quality of study of students operates [13].

*B. Modular and rating system*

The task of modularly-rating system consists in support of a complex estimation of a level of mastering by students of the main educational programs of the higher vocational training. Mastering of everyone studied by the student in a subject matter semester, including optional and elective courses, it is estimated from 0 to 100 rating points («100 % of success»). At university the following scale of recalculation of a total amount of rating points in customary estimations on four-point system operates: less than 55 - «2», from 55 to 69 - «3», from 70 to 84 - «4», from 85 to 100 - «5». Formation of a rating point in flow of a semester depending on wishes of chair and specificity of a subject matter probably in two variants resulted in the Table I [13].

TABLE I. OPTIONS OF DISTRIBUTION RATING POINTS

| № | Indicator | Option 1 | Option 2 |
|---|---|---|---|
| 1 | The maximum sum of rating points by results of progress | 60 | 80 |
| 2 | The maximum sum of rating points at examination or offset | 40 | 20 |
| 3 | Amount of the points received in a semester since which bonus | 50 | 70 |
| 4 | The maximum quantity of bonus points for successful | 30 | 20 |
| 5 | Amount of points necessary for admission of the student to | 35 | 45 |

*C. Current monitoring*

Monitoring of progress of students is subdivided on leaking, intermediate and total. Current monitoring concern: control of attendance by students of occupations, their knowledge and skills on occupations, executions of examinations; testing on sections of subject matters and so forth [13].

*D. The intermediate certification*

The intermediate certification of students is a following type of the control actions, based on results of current monitoring. It serves for an estimation of volume and level of mastering as students of a teaching material of one unit of discipline [13].

*E. Total certification*

Total certification of students is based on results of the intermediate monitoring, is led in the form of offsets, examinations, protection of term papers or projects, reports on practice and serves for an estimation of knowledge, skills of the student on all volume of a subject matter studied in a semester [13].

*F. The Pivot table and the schedule of the academic debts*

The automated workplace for a teaching department allows to carry out monitoring of progress of students. For this purpose there are following possibilities [13]:

- The Pivot table of the academic debts on semester with the shared information about the student (a surname, a name, course, group, a floor, mode of study and a mean score) and date of the last delivery. For convenience of operation probably to produce sorting, editing and a filtration of the data. It is shown on Fig. 2.
- The Schedule of the academic debts allows to display an integral sight at progress more visually. It is shown on Fig. 7.

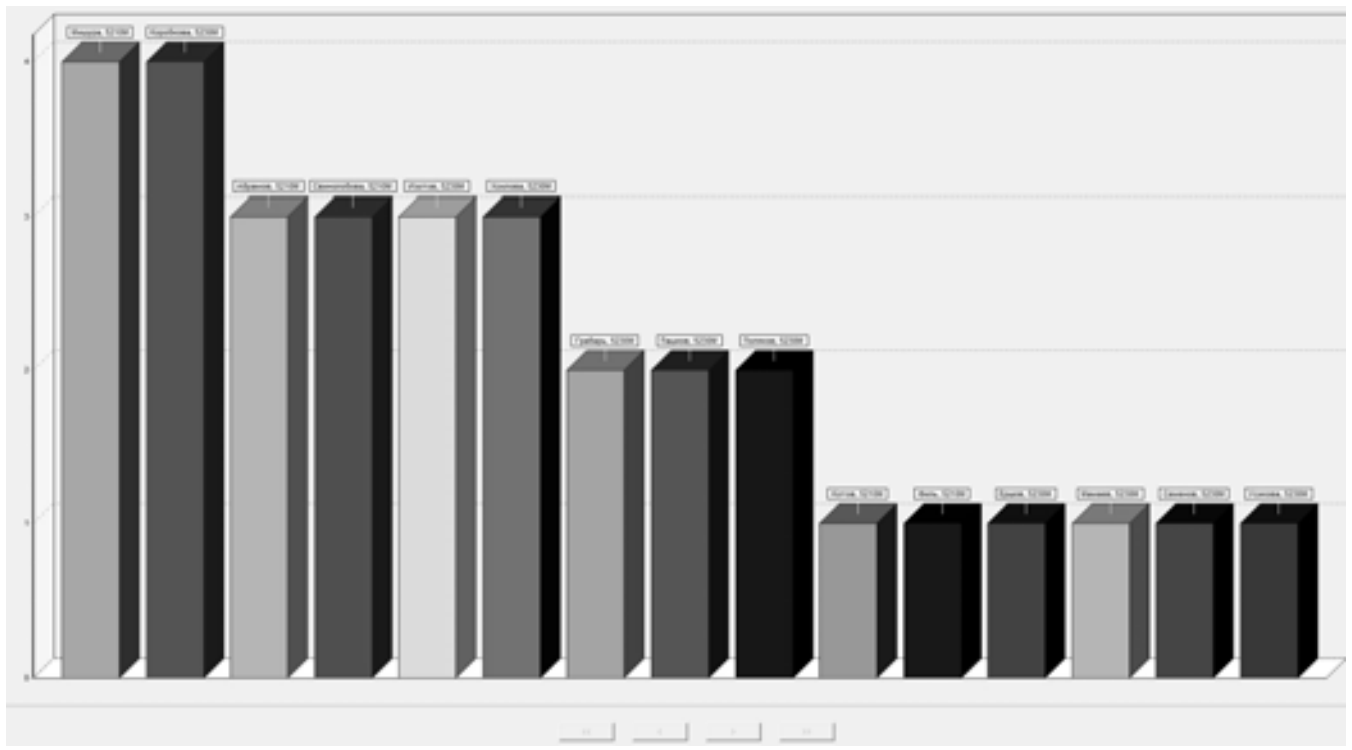

Fig. 7. The Schedule of the academic debts

## IV. COMPILATION OF THE SCHEDULE ADDITIONAL SESSION

Monitoring mastering of subject matters by students is necessary for drawing up of the schedule of additional session. During additional session the repeating examinations the students who have received unsatisfactory estimates in winter session, will be organized in winter vacation, and received an unsatisfactory estimates in spring session within one week right after the completion of session or within one week previous the beginning of new academic year. Chairs allow students to such examinations in the direction of the dean of faculty on the basis of the personal statement of the student.

Monitoring mastering of subject matters by students is shown on Fig. 8. The disciplines not handed over more than 30% of students are marked by the red. The disciplines which have been handed over by 100% of students, are marked by the green. Remaining disciplines - yellow. The table can be sorted and printed.

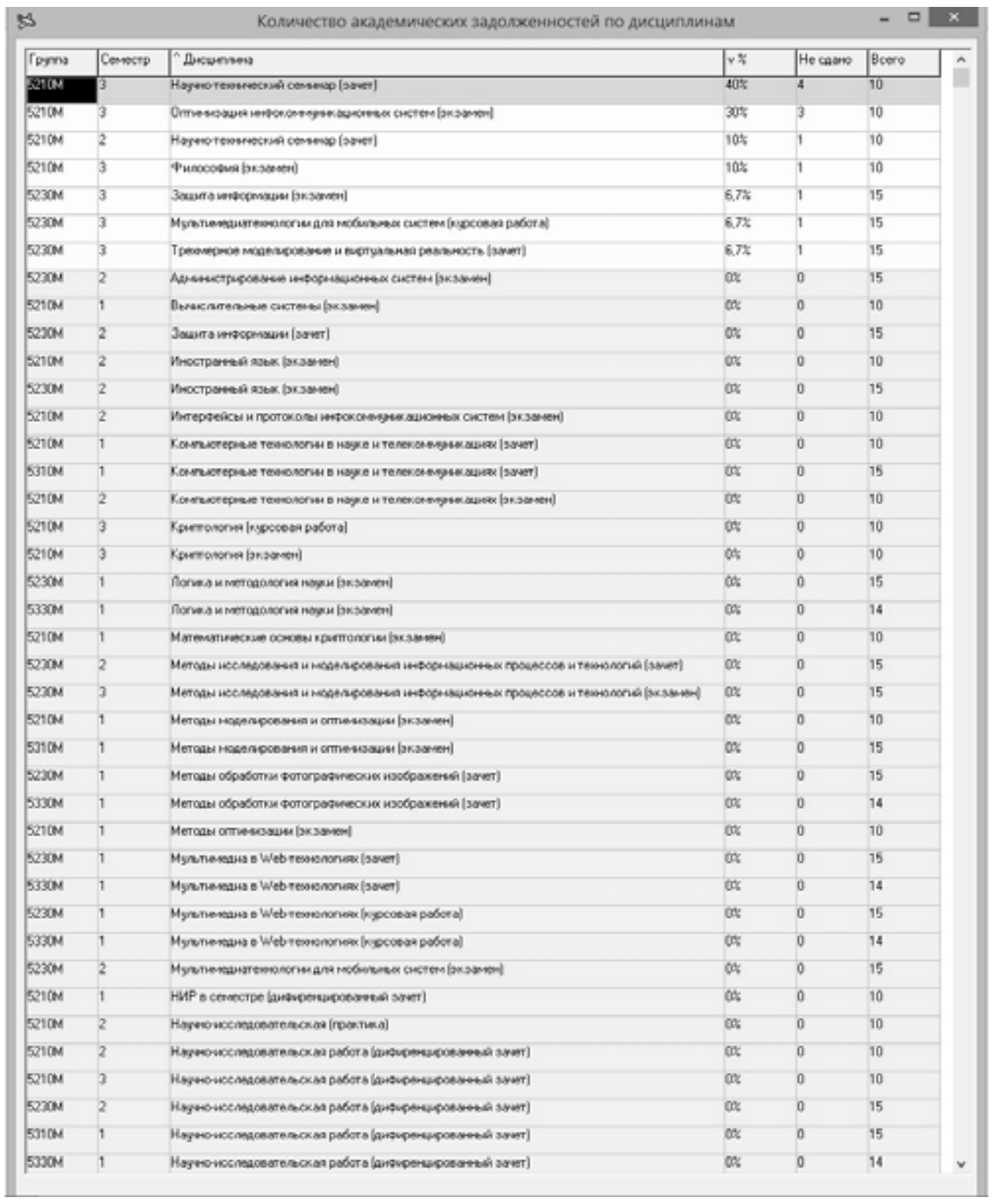
Количество академических задолженностей по дисциплинам

| Группа | Семестр | ^ Дисциплина | v % | Не сдано | Всего |
|---|---|---|---|---|---|
| 5210M | 3 | Научно-технический семинар (зачет) | 40% | 4 | 10 |
| 5210M | 3 | Оптимизация инфокоммуникационных систем (экзамен) | 30% | 3 | 10 |
| 5210M | 2 | Научно-технический семинар (зачет) | 10% | 1 | 10 |
| 5210M | 3 | Философия (экзамен) | 10% | 1 | 10 |
| 5230M | 3 | Защита информации (экзамен) | 6,7% | 1 | 15 |
| 5230M | 3 | Мультимедиатехнологии для мобильных систем (курсовая работа) | 6,7% | 1 | 15 |
| 5230M | 3 | Трехмерное моделирование и виртуальная реальность (зачет) | 6,7% | 1 | 15 |
| 5230M | 2 | Администрирование информационных систем (экзамен) | 0% | 0 | 15 |
| 5210M | 1 | Вычислительные системы (экзамен) | 0% | 0 | 10 |
| 5230M | 2 | Защита информации (зачет) | 0% | 0 | 15 |
| 5210M | 2 | Иностранный язык (экзамен) | 0% | 0 | 10 |
| 5230M | 2 | Иностранный язык (экзамен) | 0% | 0 | 15 |
| 5210M | 2 | Интерфейсы и протоколы инфокоммуникационных систем (экзамен) | 0% | 0 | 10 |
| 5210M | 1 | Компьютерные технологии в науке и телекоммуникациях (зачет) | 0% | 0 | 10 |
| 5310M | 1 | Компьютерные технологии в науке и телекоммуникациях (зачет) | 0% | 0 | 15 |
| 5210M | 2 | Компьютерные технологии в науке и телекоммуникациях (экзамен) | 0% | 0 | 10 |
| 5210M | 3 | Криптология (курсовая работа) | 0% | 0 | 10 |
| 5210M | 3 | Криптология (экзамен) | 0% | 0 | 10 |
| 5230M | 1 | Логика и методология науки (экзамен) | 0% | 0 | 15 |
| 5330M | 1 | Логика и методология науки (экзамен) | 0% | 0 | 14 |
| 5210M | 1 | Математические основы криптологии (экзамен) | 0% | 0 | 10 |
| 5230M | 2 | Методы исследования и моделирования информационных процессов и технологий (зачет) | 0% | 0 | 15 |
| 5230M | 3 | Методы исследования и моделирования информационных процессов и технологий (экзамен) | 0% | 0 | 15 |
| 5210M | 1 | Методы моделирования и оптимизации (экзамен) | 0% | 0 | 10 |
| 5310M | 1 | Методы моделирования и оптимизации (экзамен) | 0% | 0 | 15 |
| 5230M | 1 | Методы обработки фотографических изображений (зачет) | 0% | 0 | 15 |
| 5330M | 1 | Методы обработки фотографических изображений (зачет) | 0% | 0 | 14 |
| 5210M | 1 | Методы оптимизации (экзамен) | 0% | 0 | 10 |
| 5230M | 1 | Мультимедиа в Web-технологиях (зачет) | 0% | 0 | 15 |
| 5330M | 1 | Мультимедиа в Web-технологиях (зачет) | 0% | 0 | 14 |
| 5230M | 1 | Мультимедиа в Web-технологиях (курсовая работа) | 0% | 0 | 15 |
| 5330M | 1 | Мультимедиа в Web-технологиях (курсовая работа) | 0% | 0 | 14 |
| 5230M | 2 | Мультимедиатехнологии для мобильных систем (экзамен) | 0% | 0 | 15 |
| 5210M | 1 | НИР в семестре (дифференцированный зачет) | 0% | 0 | 10 |
| 5210M | 2 | Научно-исследовательская (практика) | 0% | 0 | 10 |
| 5210M | 2 | Научно-исследовательская работа (дифференцированный зачет) | 0% | 0 | 10 |
| 5210M | 3 | Научно-исследовательская работа (дифференцированный зачет) | 0% | 0 | 10 |
| 5230M | 2 | Научно-исследовательская работа (дифференцированный зачет) | 0% | 0 | 15 |
| 5310M | 1 | Научно-исследовательская работа (дифференцированный зачет) | 0% | 0 | 15 |
| 5330M | 1 | Научно-исследовательская работа (дифференцированный зачет) | 0% | 0 | 14 |

Fig. 8. The Schedule of the academic debts

## V. EDITING INFORMATION ABOUT STUDENTS

In our application it is possible to carry out editing dates of deposit disciplines and personal information on students. The window of the first editing shares on 3 logic parts (fig. 5):

- Preservation and exit. There are two buttons: "Exit" - closes a window without preservation, and «To keep and close a window» - keeps changes and closes a window.
- Date and operating mode choice. Consists of a calendar and the menu of a mode of editing. In the first - date which will be added in the chosen cell gets out. The chosen date remains, that is at repeated opening of a window the last chosen date will be active. In the second - the editing mode gets out: "viewing" - at a cell choice it will be allocated, "insert" - will be inserted the date chosen in a calendar, to "remove" - the chosen cell will be cleared. The calendar is active only in the insert mode.
- Editing. It is the table where dates of deposit disciplines for a semester are displayed. The filled cells are allocated green, and empty - red. Editing can be carried out on three categories: on a semester (Fig. 9), on the student (Fig. 10) both the chosen student and a semester (Fig. 11).

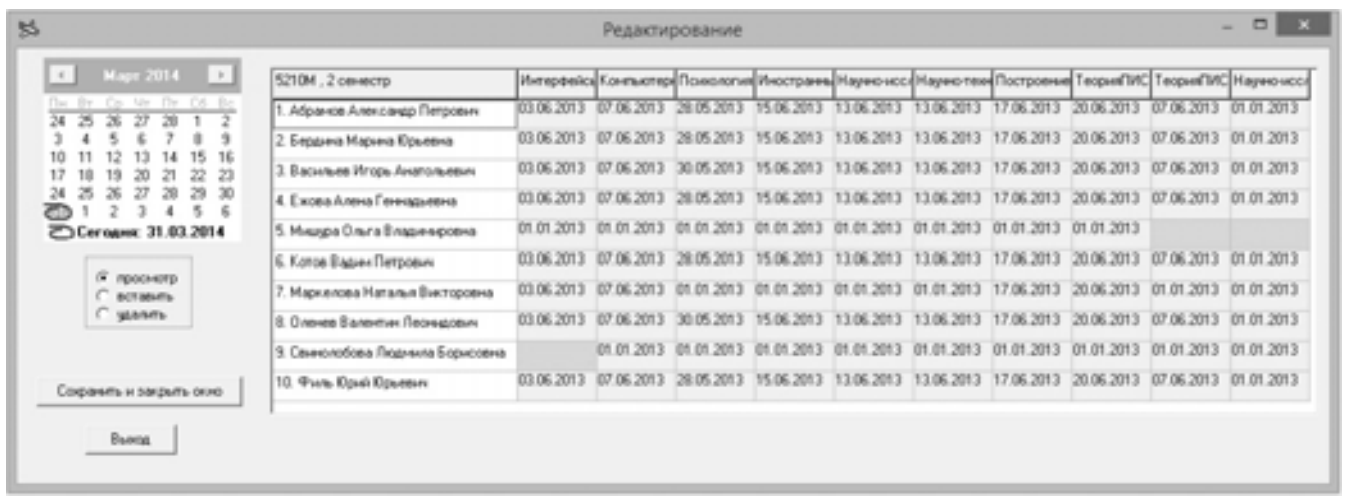

Fig. 9. Editing the academic debts for the semester

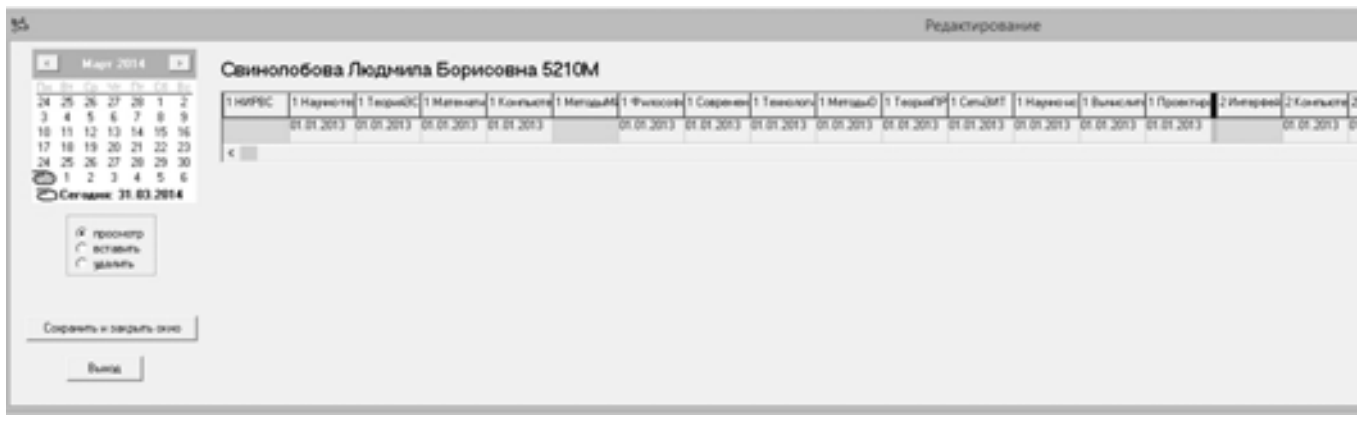

Fig. 10. Editing the academic debts for the student

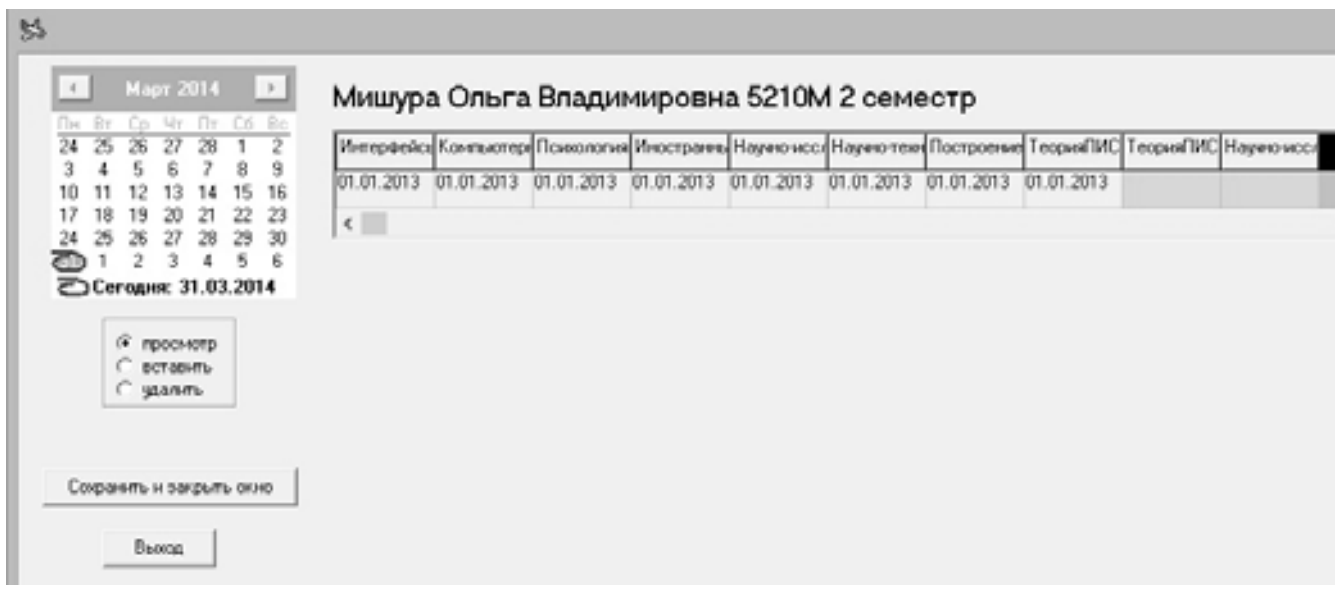

Fig. 11. Editing the academic debts for the semester and student

The second allows to edit such data, as a surname, a name, a patronymic, student ID card number, a GPA and a floor. Besides, in browse mode it is possible to see information on the training direction, group, a course, form of education and history of orders on the student. It is shown on Fig. 12.

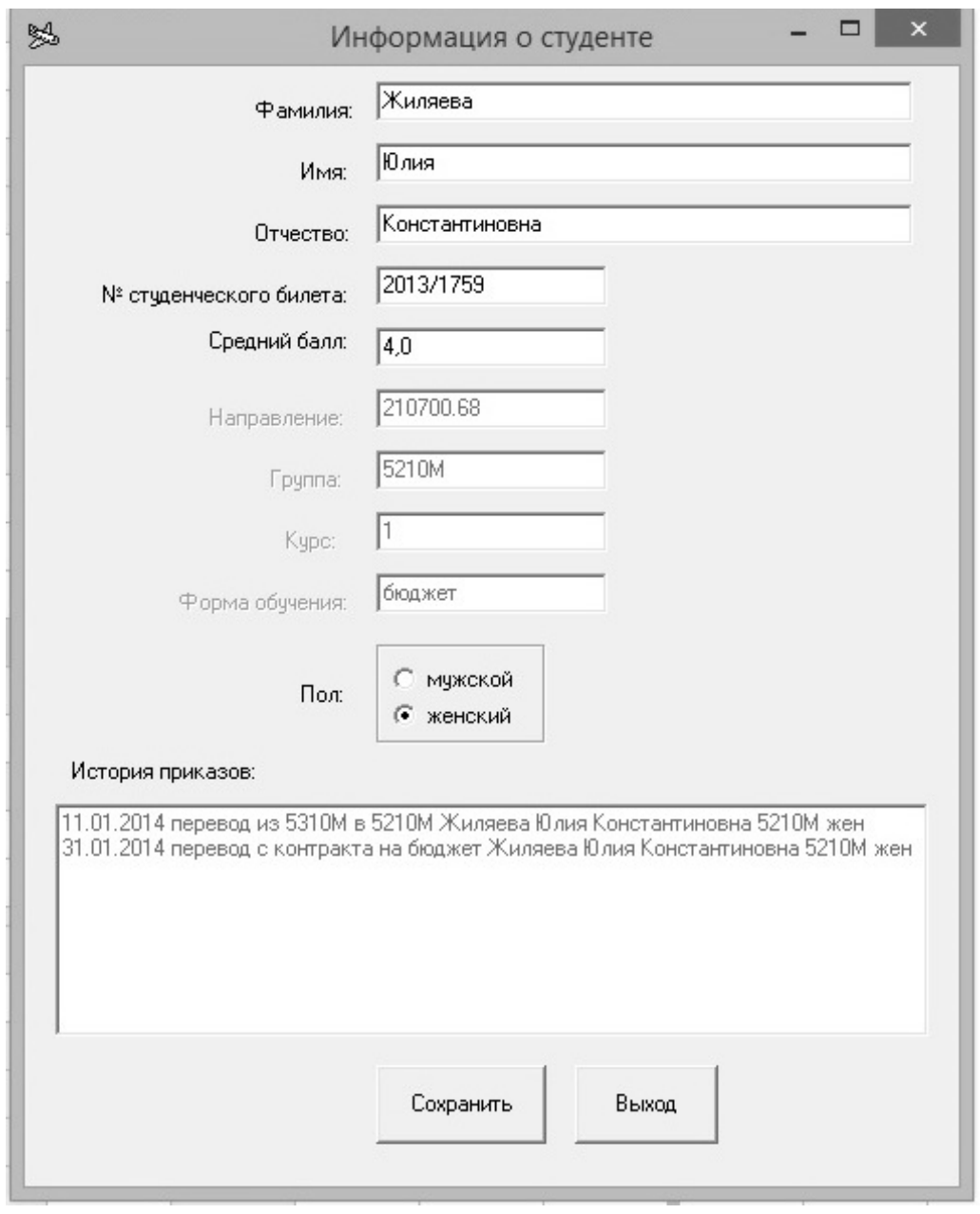

Fig. 12.Edit the personal informational student

## VI. Driving of a contingent of students

### *A. Driving of a contingent of students*

Movement contingent students - assignment, enrollment and transfers of students. Example of this operations is shown on Fig 13 and 14. It is recorded in the file report.log and shown on Fig 15. The report can be printed out.

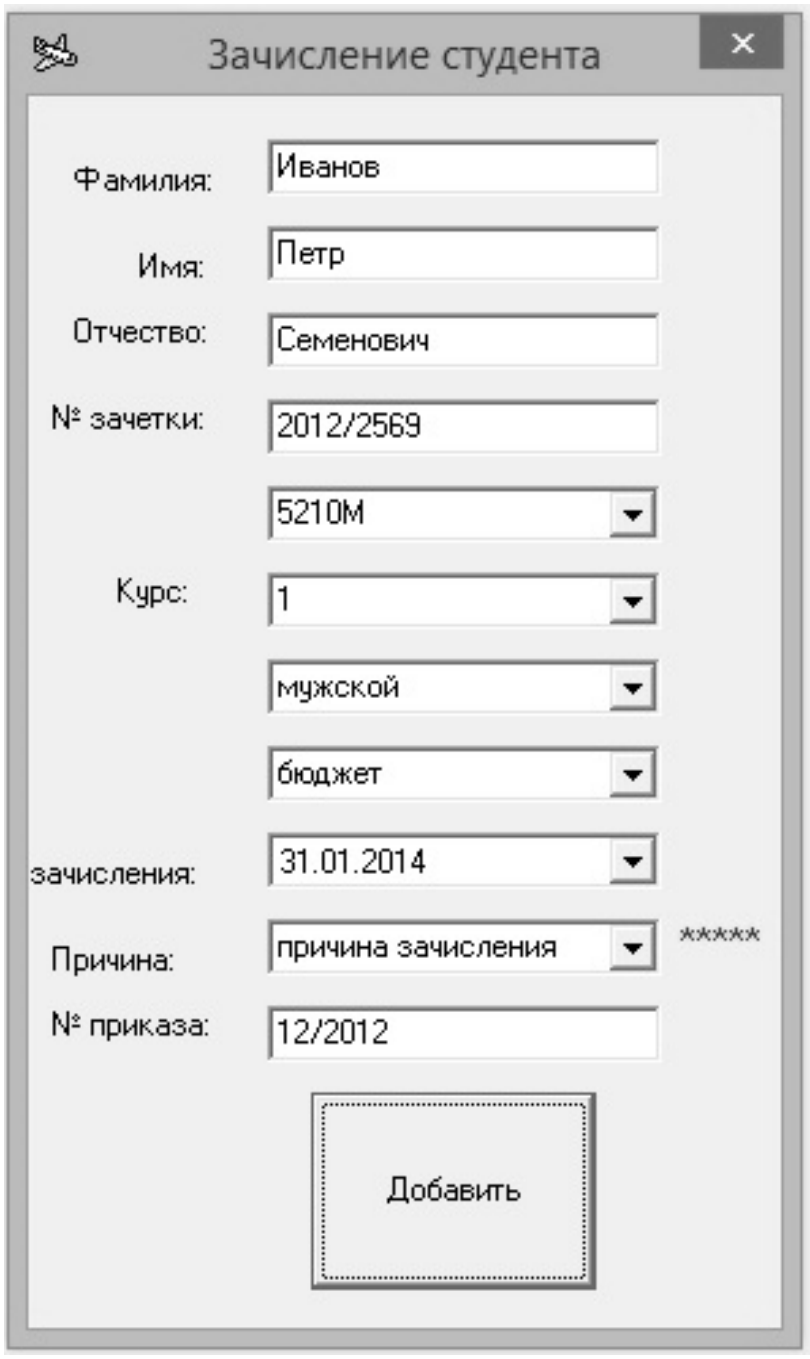

Fig. 13. Assignment student

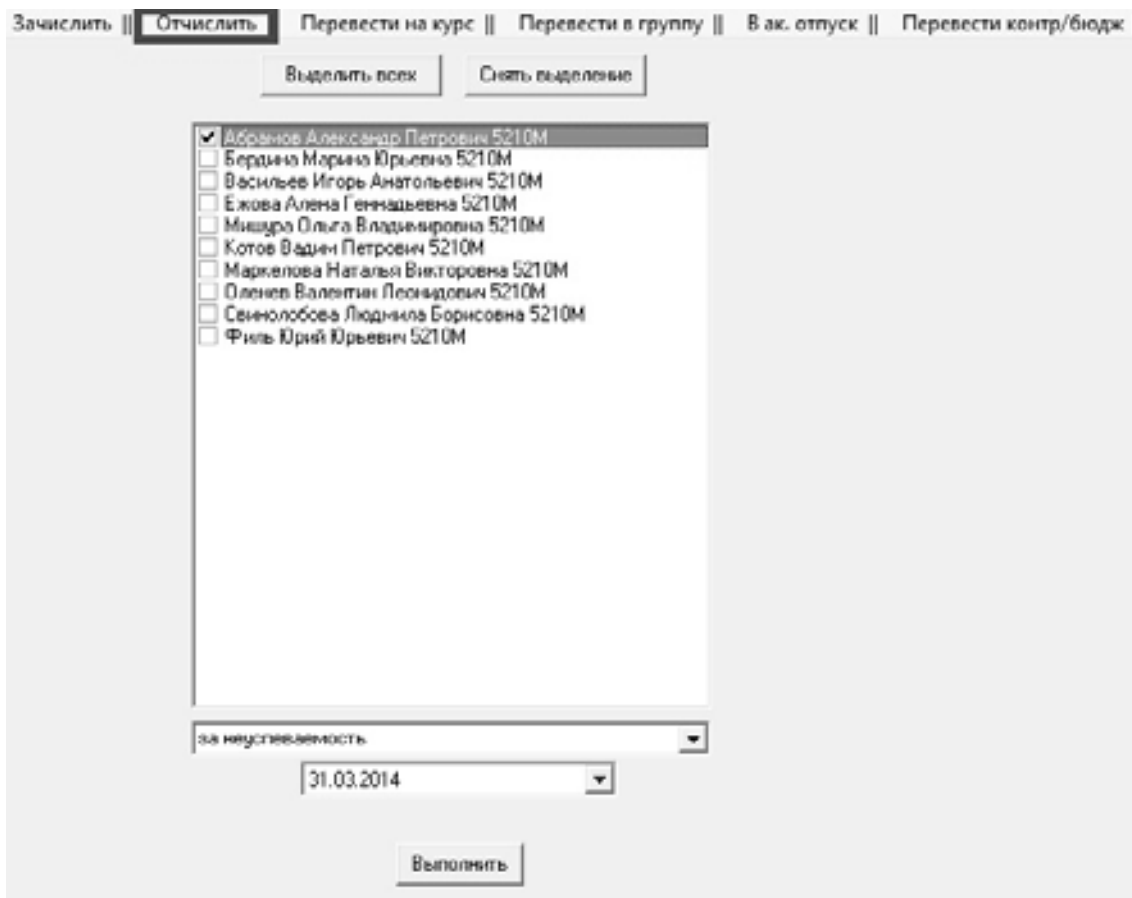

Fig. 14. Enrollment students

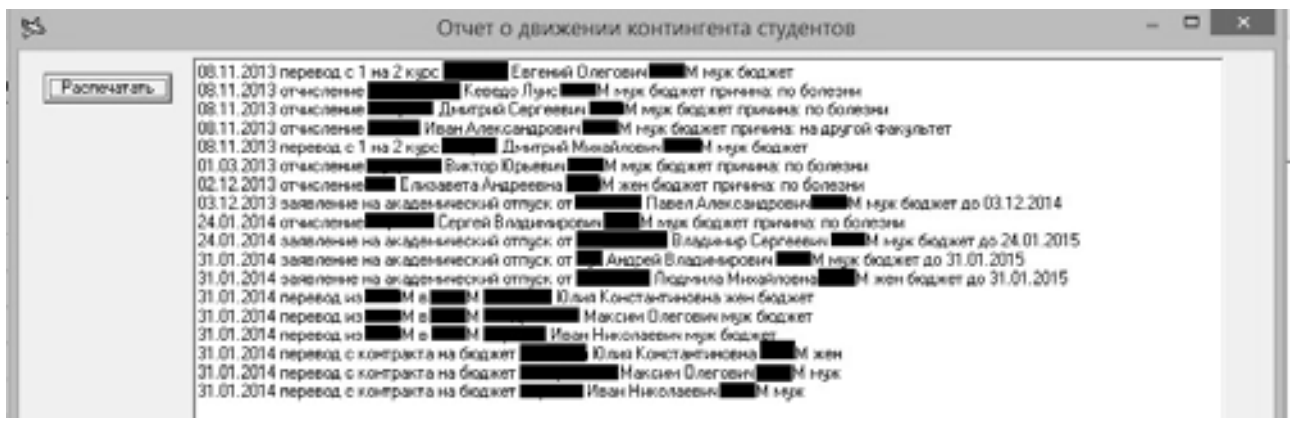

Fig. 15. Report movement contingent students

### *B. The report movement contingent students*

The automated workplace for a teaching department automatically forms statistical reports of a different look. The statistical report is an official document which contains data on work of the accountable object, brought on a special form. One of such reports is the report movement contingent students (Fig. 16) is a report which shows amount arrived, left, and the transferred students, considering contract/budgetary forms of education and a floor (man's/female). It is made for the periodic reporting of higher education institution before the Ministry of Education and Science of the Russian Federation and unloaded in Microsoft Excel [12].

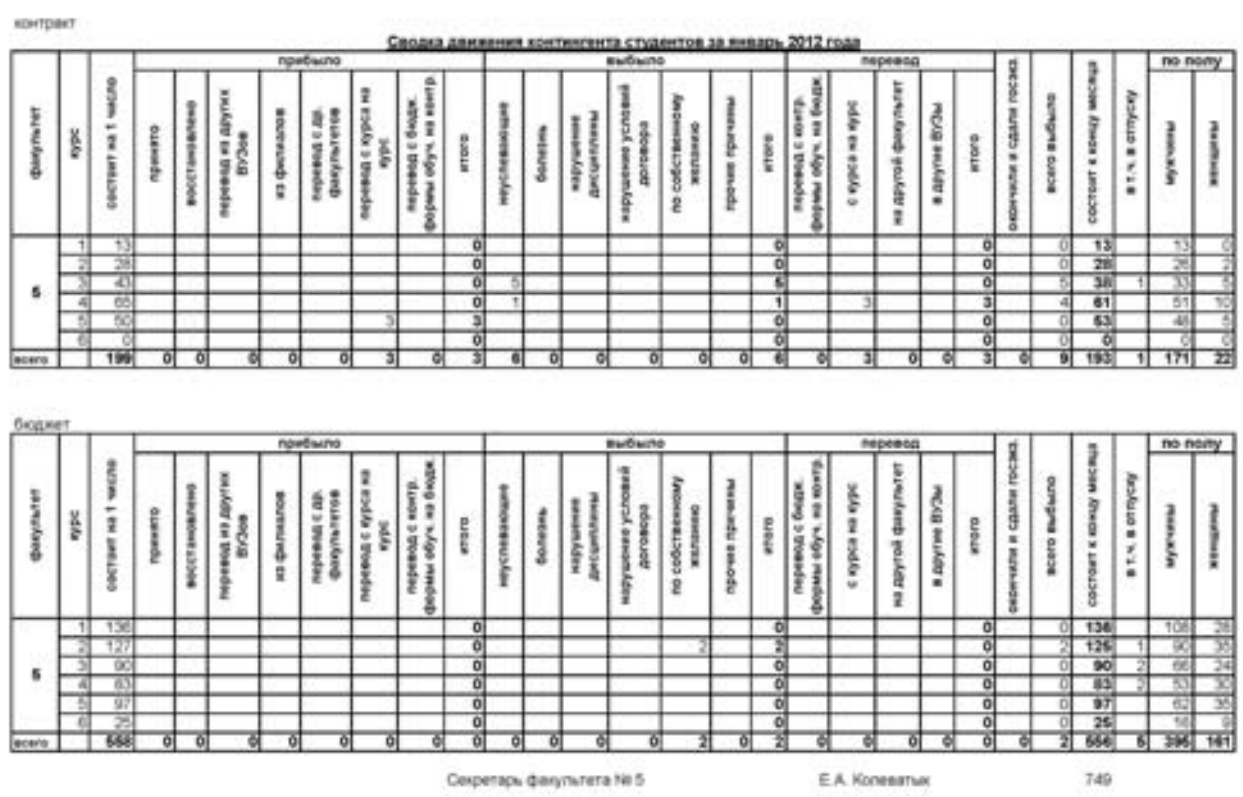

Fig. 16. Report movement contingent students

## VII. CONNECTION WITH AN AUTOMATED WORKPLACE OF THE HIGHER SCHOOL TEACHER

As it was planned in [1] development of an automated workplace of the teacher of the higher school [14] is made. In this chapter I will consider the problem connection of its with an automated workplace of a teaching department and briefly I will tell about it.

The idea creation automated workplace the teacher arose on purpose to simplify work teachers, to release their hands from primitive teaching magazines, to accelerate process of a gain score students, to reduce probability a human mistake at calculation points, to provide impossibility change data with the third parties, to eliminate financial costs of the printing of magazines and to convict distribution of information on progress of students [15].

The tasks that it will implement:

- implementation of electronic log which will integrate in itself attendance log, log with estimates, teaching log and log according to the security regulation,
- data collection automation about the current, intermediate and total certifications,
- data collection automation about attendance,
- automation the process of assessment,
- sheet filling. It is shown on Fig. 17.

**МИНИСТЕРСТВО ОБРАЗОВАНИЯ И НАУКИ РОССИЙСКОЙ ФЕДЕРАЦИИ**
Федеральное государственное автономное образовательное учреждение высшего профессионального образования
'Санкт-Петербургский государственный университет аэрокосмического приборостроения'

**ВЕДОМОСТЬ УЧЕТА ТЕКУЩЕЙ УСПЕВАЕМОСТИ 2013/2014 уч.год**

Институт 5 Специальность 23040062 Семестр 5 №гр 5131
Кафедра 53

Дисциплина Кроссплатформенное програмирование

| Дата 13.01.14 | | | ЭКЗАМЕН | | | | |
|---|---|---|---|---|---|---|---|
| Преподаватель Иванов И.В. | | | | | | | |
| № | Фамилия И.О. | № зач. книжки | Рейтинг семестра | Итог. рейтинг | Оценка | Дата | Подпись |
| 1 | Абрамов А.В. | 2011/0856 | 45 | 65 | удовл | 13.01.14 | |
| 2 | Бердина Д.В. | 2011/0373 | | | неявка | | |
| 3 | Васильев А.М. | 2011/0376 | | | неявка | | |
| 4 | Вересов И.У. | 2011/0378 | 60 | 89 | отлично | 13.01.14 | |
| 5 | Грабарь С.А. | 2011/0381 | | | неявка | | |
| 6 | Ежова И.О. | 2011/0385 | 48 | 70 | хорошо | 13.01.14 | |
| 7 | Ершов К.В. | 2011/0387 | 45 | 70 | хорошо | 13.01.14 | |
| 8 | Изотов А.Е. | 2011/0388 | 60 | 100 | отлично | 13.01.14 | |
| 9 | Коробков Н.А. | 2011/0389 | 45 | 65 | удовл | 13.01.14 | |
| 10 | Кулин К.В. | 2010/0317 | 55 | 85 | отлично | 13.01.14 | |
| 11 | Мамаев И.С. | 2011/0289 | | | неявка | | |
| 12 | Михалев В.А. | 2011/0411 | 49 | 70 | хорошо | 13.01.14 | |
| 13 | Мишура Е.А. | 2011/0412 | | | неявка | | |
| 14 | Оленев И.В. | 2011/0421 | 55 | 70 | хорошо | 13.01.14 | |
| 15 | Пашков А.И. | 2011/1235 | 41 | 70 | хорошо | 13.01.14 | |
| 16 | Поляков В.В. | 2011/1237 | 52 | 91 | отлично | 13.01.14 | |
| 17 | Свинопобова Р.О. | 2011/0428 | 40 | 60 | удовл | 13.01.14 | |
| 18 | Семенов Д.И. | 2011/0429 | 45 | 65 | удовл | 13.01.14 | |
| 19 | Сыщиков В.М. | 2011/0431 | 60 | 100 | отлично | 13.01.14 | |
| 20 | Усикова В.А. | 2011/1267 | 60 | 100 | отлично | 13.01.14 | |
| 21 | Филь А.Д. | 2011/0437 | 58 | 85 | отлично | 13.01.14 | |
| 22 | Хохлов К.С. | 2011/0443 | 35 | 60 | удовл | 13.01.14 | |

ИТОГО: Отлично____7____
Хорошо_____5____
Удовлетв.____5____
Неудовлетв.___0___
Неявка_____5____

Директор института__________________ Е.А. Крук

Fig. 17. Examination list

Besides, this application works considering modular and rating system. I told about it in the section III.B, but here I will tell in more detail.

Features of modular and rating system of GUAP, which are considered by an automated workplace of the teacher:

- Duration of theoretical training in a semester makes 17 weeks, examinations duration – 3,5 weeks.
- Faultless assimilation of everyone studied by the student in a semester of a subject matter is estimated at 100 rating points.
- For the disciplines which study is carried out some semester, 100 points are selected in each semester.
- Course design (term paper) are evaluated on discipline separately, thus the maximum number of points also is set equal 100.
- When charging points for execution of laboratory operations and other jobs it is expedient to consider timeliness of their execution.
- From the maximum number of points which the student can gain in a semester, a quantity of points should be led out for visit of occupations and the academic activity of students.
- The repeating examinations the students who have received unsatisfactory estimates in winter session, will be organized in the period of winter vacation, and received unsatisfactory estimates in spring session, within one week right after the end of session or within one week previous the beginning of new academic year. Chairs allow students to such examinations in the direction of the dean of faculty based on personal announcement of the student.

The application is written to Delphi 7 in the Object Pascal language [5-7,16].

Import and export to an automated workplace of a teaching department happens automatically.

## VIII. SYSTEM AUTOMATIC SEARCH MISTAKES

The system of automatic search of mistakes is developed for prevention emergence mistakes in data of the program: timeliness of an exit from the academic holiday, the completion of training and transition to the following course. At first sight it seems that such mistakes never will appear, but it is true for a case when at you 10 or 20 students and when their number exceeds 500, this system becomes actual. The report of this system is shown on Fig. 18.

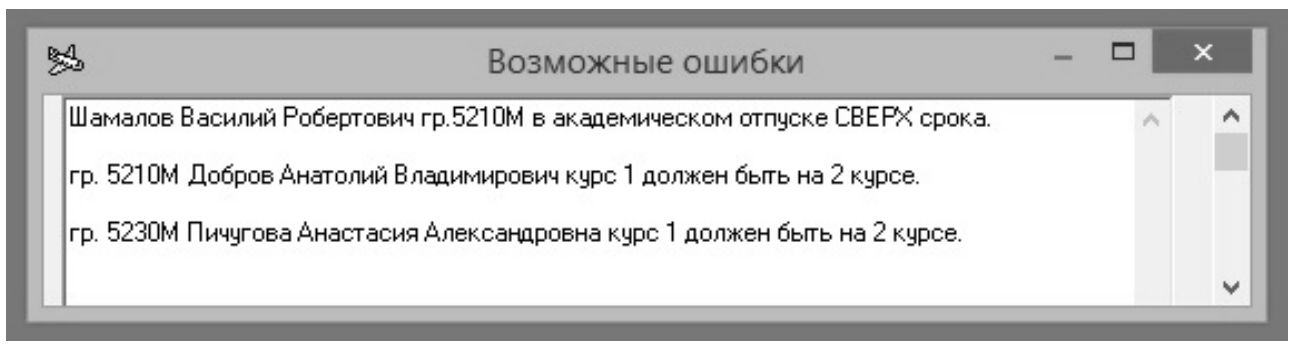

Fig. 18. System automatic search mistakes

## IX. CONCLUSION

I developed a program part of an automated workplace of a teaching department which allows to carry out monitoring progress students, monitoring mastering of subject matters by students, synchronization with an automated workplace of the higher school teacher and automatically fills the report of movement of the contingent of students.

Moreover, the designed system allows to increase the efficiency and effectiveness of staff training center [1].

## ACKNOWLEDGMENT

I would like to thank my scientific head - Alexey Zhukov for his pedagogical and scientific approach, excellent knowledge of a subject and sincere fatherly support which have inspired me on performance of the given work. His recommendations and councils helped me for all runtime of operation.

Hearty thanks to my family for support and the help.

## REFERENCES

[1] V.A. Ushakov, "*Development of dean work area*", *XVII Reshetnevsky Readings*, Nov. 2013, pp. 263-265.

[2] V.A. Ushakov, I. U. Bkhattachardzhi "*Implementation hardware at automated workplace for a teaching department*", unpublished

[3] V.A.Ushakov, "*Automated workplace for a teaching department*", *66th GUAP Conf.*, Apr. 2013, pp. 342-346.

[4] V.V. Faronov, *Delphi. Programming in high level languages.* Saint-Petersburg, Lider, 2010, 640p.

[5] A.D. Khomonenko, *Delphi 7,* Saint-Petersburg, BHV-Petersburg, 2008, 1216 p.

[6] A.IA. Arkhangelskii, *100 delphi components*, Web: http://www.beluch.ru/progr/100comp/index.htm.

[7] *Delphi bases*, Web: http://www.delphibasics.ru/.

[8] K. Pachenko and S. Teykseyra, *Delphi5. Developer's guide.* In two volumes, Volume one, Moscow, Williams, 2000, 832 p.

[9] V.N. Kornyakov *Programming documents and applications of MS Office in Delphi.* Saint-Petersburg, BHV-Petersburg, 2006, 496 p.

[10] *Fast data processing of Excel in Delphi*, We program in Delphi, Web: http://www.webdelphi.ru/2012/01/bystraya-obrabotka-dannyx-excel-v-delphi/

[11] A. Garnaev and L. Rudikova, *Microsoft Office Excel 2010: Applications programming,* Saint-Petersburg, BHV-Petersburg, 2011, 514 p.

[12] IU. Magda, *Applications programming of Microsoft Office 2007 in Delphi,* Saint-Petersburg, BHV-Petersburg, 2009, 160 p.

[13] V.A. Ushakov, I. U. Bkhattachardzhi "*Monitoring progress students with use of an automated workplace for a teaching department*", unpublished.

[14] I. U. Bkhattachardzhi, V.A. Ushakov "*Automated workplace of the teacher of the higher school*", unpublished.

[15] A.N. Lebedev and S.V. Kotelnikova, "*Development of electronic register for a teacher with rating system of student grades*", *XVI Reshetnevsky Readings*, Nov. 2012, pp. 576-577.

[16] M. Kentu, *Delphi 7: for professionals,* Saint-Petersburg, Piter, 2004, 1101 p.